\documentclass[10pt,a4paper,twoside,reqno]{amsart}
\usepackage[cp1251]{inputenc}
\usepackage[russian,english]{babel}
\usepackage{amsmath,amsthm,amssymb,graphicx,color}
\usepackage[colorlinks,pdfauthor={Yu. Brezhnev},
            bookmarksopen=false,bookmarks=false]{hyperref}

\def~{\unskip\nobreak\hspace{0.27778em}\ignorespaces}
\def\,{\ifmmode\mskip+1.5mu\else\kern+.08333em\fi\relax}
\def\!{\ifmmode\mskip-1.5mu\else\kern-.08333em\fi\relax}
\newdimen{\FontSize} \delimiterfactor=990

\def\ds{\displaystyle}  \def\sss{\scriptscriptstyle}
\def\ie{i.\,\,e.}  \def\eg{e.\,\,g.}
\def\re{{\,\mathrm{e}}} \def\ri{{\mathrm{i}}}

\def\ep{e\hbox{\smaller[1]$'$}}
\def\epp{e\hbox{\smaller[1]$'\!'$}}

\def\DEF{\mathrel{\vcenter{\hbox{$:$}}{=}}}

\def\hence{=\kern-0.5em\Rightarrow}
\def\?{\textrm{\protect\footnotesize$\RED\mathchar"446$}}
\def\vphi{\mskip0mu\raisebox{-0.54ex}{\scalebox{0.6}[1]{\hbox{$-$}}}
 \mkern-7.76mu\raise0.52ex\hbox{$\varphi$}}
\def\vpsi{\mskip0.4mu\raisebox{-0.54ex}{\scalebox{0.6}[1]{\hbox{$-$}}}
 \mkern-8.82mu\raise0.52ex\hbox{$\psi$}{}}

\def\russian{\selectlanguage{russian}}

\def\END{\end{document}}

\def\Mfrac#1#2{\def\FRAC{\hbox{\smaller[1]$\ds\frac{#1}{#2}$}}
\ifdim\FontSize=11pt\raise0.051ex\FRAC\else\raise0.059ex\FRAC\fi}
\def\mfrac#1#2{\def\FRAC{\hbox{\smaller[3]$\ds\frac{#1}{#2}$}}
\ifdim\FontSize=11pt\raise0.13ex\FRAC\else\raise0.174ex\FRAC\fi}
\newcommand{\AB}[2]{\nobreak\mskip-1mu
\raise0.12ex\hbox{\smaller[2]\renewcommand{\arraystretch}{0.42}%
\setlength{\arraycolsep}{0pt}\raise0.16ex\hbox{$\big[$}%
$\begin{array}{c}\scriptstyle#1\\ \scriptstyle#2\end{array}$%
\raise0.16ex\hbox{$\big]$}}\mskip-1mu\relax}

\newcommand{\mbig}[2][4]{\vcenter{\hbox{%
\ifcase#1$#2$\or\small$\big#2$\or$\big#2$\or\large$\big#2$%
\or\footnotesize$\Big#2$\or\small$\Big#2$\or$\Big#2$\or\large$\Big#2$%
\or$\bigg#2$\or\large$\bigg#2$\or$\Bigg#2$\or\large$\Bigg#2$%
\or\Large$\Bigg#2$\else\huge$\bigg#2$\fi}}\relax}

\newdimen{\upp}\newdimen{\D}\newdimen{\DD} 
\newcommand{\pow}[4][\displaystyle]{
\settowidth{\upp}{\hbox{$#1 {}^{#4}$}}%
\settowidth{\D}{\hbox{$#1{#2}$}}%
\settowidth{\DD}{\hbox{$#1{#2}_{#3}$}}%
\addtolength{\DD}{-\D} {#1{#2}_{#3}\kern-\DD {}^{#4}}
\addtolength{\DD}{-\upp}\ifdim\DD>0pt \kern+\DD \fi\relax}

\makeatletter 
\providecommand{\larger}[1][1]{
\ifmmode\relax\else\@tempcnta\ifx\@currsize\normalsize 4\else
 \ifx\@currsize\small 3\else\ifx\@currsize\footnotesize 2\else
 \ifx\@currsize\large 5\else\ifx\@currsize\Large 6\else
 \ifx\@currsize\LARGE 7\else\ifx\@currsize\scriptsize 1\else
 \ifx\@currsize\tiny  0\else\ifx\@currsize\huge 8\else
 \ifx\@currsize\Huge  9\else 4 \fi\fi\fi\fi\fi\fi\fi\fi\fi\fi
\advance\@tempcnta#1\relax\ifnum\@tempcnta<\z@ \@tempcnta\z@ \fi
\ifcase\@tempcnta\tiny\or\scriptsize\or\footnotesize\or\small
 \or\normalsize\or\large\or\Large\or\LARGE\or\huge\else\Huge\fi\fi}

\providecommand{\smaller}[1][1]{\larger[-#1]} \makeatother

\makeatletter \def\@dotsep{1} \makeatother 

\def\RED{\color{red}} \def\BLUE{\color{blue}} 

\newcommand{\Width}[2][\displaystyle]
{\settowidth\D{\hbox{$#1#2$}}\texttt{\tiny\BLUE$($\the\D$)$}}
\newcommand{\Height}[2][\displaystyle]
{\settoheight\D{\hbox{$#1#2$}}\texttt{\tiny\BLUE$($\the\D$)$}}
\newcommand{\Depth}[2][\displaystyle]
{\settodepth\D{\hbox{$#1#2$}}\texttt{\tiny\BLUE$($\the\D$)$}}

\usepackage{hyperref}

\newcommand{\textcenter}[3][Yu.~Brezhnev]{
\pagestyle{myheadings}
\markboth{\hfill\textsc{#1}\hfill}{\hfill\textsc{#1}\hfill}
\ifdim#2mm=0mm\textwidth=\textwidth\else\textwidth=#2mm\fi
\oddsidemargin=\paperwidth%
\addtolength{\oddsidemargin}{-\textwidth}%
\addtolength{\oddsidemargin}{-2in} \oddsidemargin=0.5\oddsidemargin
\evensidemargin=\oddsidemargin
\ifdim#3mm=0mm\textheight=\textheight\else\textheight=#3mm\fi
\topmargin=\paperheight%
\addtolength{\topmargin}{-\textheight}%
\addtolength{\topmargin}{-2in}%
\addtolength{\topmargin}{-\headsep}%
\addtolength{\topmargin}{-\headheight}%
\addtolength{\topmargin}{-\footskip}%
\topmargin=0.5\topmargin}

\newcommand\obrezka[2]{
\oddsidemargin=#1mm \paperwidth=\textwidth%
\addtolength{\paperwidth}{2\oddsidemargin}%
\addtolength{\oddsidemargin}{-1in} \evensidemargin=\oddsidemargin%
\topmargin=#2mm \paperheight=\textheight%
\addtolength{\paperheight}{\headsep}%
\addtolength{\paperheight}{\headheight}%
\addtolength{\paperheight}{\footskip}%
\addtolength{\paperheight}{2\topmargin}%
\addtolength{\topmargin}{-1in}}

 \textcenter{136}{200} \obrezka{5}{1}

\theoremstyle{plain}
\newtheorem{theorem}{Theorem}
\newtheorem{proposition}[theorem]{Proposition}

\theoremstyle{remark}
\newtheorem{remark}{Remark}
\allowdisplaybreaks[4]

\begin{document}

\begin{abstract}
We show that formulas differing from classical analogues of rational
trace formulas for algebraic-geometric potentials occur in the theory
of finite-gap integration of spectral equations. The new formulas
contain transcendental modular functions and hypergeometric series.
They result in transcendental relations for theta functions.
\end{abstract}


\author[Yu.~Brezhnev]
{Yurii V.~Brezhnev}

\title[Transcendental Trace Formulas]
{Transcendental Trace Formulas\\For Finite-Gap Potentials}

\thanks{This work was supported in part by the Russian Federal
Agency for Science and
Innovation (Grant No.~02.740.11.0238) and the Program for Supporting
Leading Scientific Schools (Grant No.~NSh-3400.2010.2).}

\hfill \noindent {\footnotesize{\sc Theor.~Math.~Phys.} {\bf 164}(1)
(2010), 920--928}
\bigskip\bigskip

\maketitle\tableofcontents\thispagestyle{empty}

\section{Intorduction}
\noindent Trace identities and formulas are known in the integration
theory of spectral problems. This terminology originates from papers by
Krein, Gel'fand, and Dickey on the spectral theory of the ordinary
differential equation of the form
$\psi_{\!\mathit{xx}}^{}-u(x)\,\psi=\lambda\,\psi$ dating back to the
1950s. In 1975, stimulated by the appearance of finite-gap integration
methods (in the spectral formulation), V.~Matveev established a new and
important interpretation of these formulas. These relations, together
with their $t$-isospectral modifications, are differential identities
for exactly solvable potentials $u(x)$. In particular, using one of the
series of these formulas, namely, the famous formula
\begin{equation}\label{1}
u(x)=2\,\big\{\gamma_1^{}(x)+\cdots+\gamma_g^{}(x)\big\}
+\mbox{const}\,,
\end{equation}
allows reconstructing the potentials from the quantities
$\gamma_k^{}(x)$, which satisfy integrable ordinary differential
equations that are the Dubrovin equations. All the remaining trace
identities follow from this ``master'' formula. Here, we interpret
these relations or, to be precise, their analogues for higher-order
spectral equations as formulas that express an exactly solvable
potential in terms of the quantities $\gamma_k^{}(x)$.

It is known that a potential  $u(x)$ determines a Liouville-integrable
finite-dimensional Hamiltonian system in which $x$ plays the role of
time. We pass to the canonical variables $(p, q)$ by combining the
potential and its derivatives: $p = p(u, u_x, \ldots)$, $q = q(u,
u_x,\ldots)$  \cite{dickey}. The subsequent transition to separation
variables is the transition to the variables  $\gamma_k^{}$  and
$\mu_k^{}$ subject to the algebraic relation
$W(\gamma_k^{},\mu_k^{})=0$, which is called the spectral curve
equation. Formulas of the inverse transition
$p=p(\gamma_k^{},\mu_k^{})$, $q=q(\gamma_k^{},\mu_k^{})$ and hence the
formula of the form $u = u(p, q)$ are well known  \cite{blaszak} in
Hamiltonian system theory, but we can clearly identify formula
\eqref{1} with a differential analogue of an operator trace only in the
Schr\"odinger equation case. A ``trace'' interpretation of all other
algebraic-geometric (finite-gap) operators is lacking. This is because
both the spectral curve coordinates $\gamma$ and $\mu$ must participate
in a general theory of variable separation because Abelian functions,
generally speaking, are expressed in terms of both these variables.
Second-order equations are the only particular case where
$u=u(\gamma)$. Only isolated examples are known for more complex
spectral problems  \cite{br}, while a general scheme for constructing
the principal trace formula
\begin{equation}\label{R}
u=R(\gamma_k^{},\mu_k^{})
\end{equation}
with a rational function $R$ is still lacking, to the best of our
knowledge.

We devote this paper not to this problem but to the observation that in
addition to representing potentials in the form of Abelian functions as
in \eqref{R}, we have other representations. They are provided by
transcendental, not rational but still \textit{single-valued},
functions of symmetric combinations of the parameters
$(\gamma_k^{},\mu_k^{})$. The example we use to demonstrate this is
interesting not only in itself but also because it leads to new
transcendental identities for $\Theta$-functions. We recall that
polynomial identities for  $\Theta$-functions are well known and are in
fact uniformizing representations for Abelian varieties  \cite{buch}
and   $\Theta$-functional representations for Abelian functions
\eqref{R} respectively.

\section{The rational trace formula}
\noindent We consider the third-order spectral problem
\begin{equation}\label{3}
\Psi'''+u(x)\,\Psi'-\frac12\,u'(x)\,\Psi=\lambda\,\Psi\,.
\end{equation}
It is known that if we associate this problem with a nonlinear
integrable equation, then its isospectral deformations are described by
the Kaup--Kupershmidt equation \cite{kaup}
\begin{equation}\label{KK}
u_t^{}=u_{\mathit{xxxxx}}+5\,u\,u_{\mathit{xxx}}+\frac{25}{2}\,u_x
u_{\mathit{xx}}+5\,u^2\,u_x\,.
\end{equation}
In what follows, it suffices to consider only stationary solutions or,
equivalently, the operator pencil
\begin{equation}\label{ABC}
A(\lambda;[u])\,\Psi''+B(\lambda;[u])\,\Psi'+
C(\lambda;[u])\,\Psi=\mu\,\Psi\,,
\end{equation}
which depends linearly on $\lambda$ and commutes with  \eqref{3}.

\subsection{The algebraic curve}
The compatibility condition of Eq.~\eqref{ABC} with the spectral
problem, \ie, with problem  \eqref{3}, customarily yields expressions
for $A$, $B$, and $C$:
$$
A=-9\,\lambda\,,\qquad B=u^2+\frac12\,u''+c\,,\qquad
C=-\frac13\,A''+\frac23\,A\,u-B'\,,
$$
The constant is related to the stationary dynamics $u = u(x - c\,t)$,
and we have
$$
-9\,\lambda\,\Psi''+\Big(u^2+\frac12\,u''+c\Big)\Psi'-
\Big(6\,u\,\lambda+2\,u\,u''+\frac12u'''\Big)\Psi=\mu\,\Psi\,.
$$
The quantities  $\lambda$ and $\mu$ are algebraically dependent, \ie,
they belong to the algebraic curve
\begin{equation}\label{W}
\mu^3+Q(\lambda;[u])\,\mu+T(\lambda;[u])=0\,.
\end{equation}
We obtain the equation of this curve by eliminating $\Psi$ from
\eqref{3} and \eqref{ABC}, with the result
$$
\mu^3+\Big(27\,c\,\lambda^2-\frac92\,K\,\lambda+E_2\Big)\mu+
729\,\lambda^5
+E_1\,\lambda^3+\frac{27}{2}\,u\,K\,\lambda^2+E_3\,\lambda+
\frac18\,(u''+2\,u^2+2\,c)^2\,K=0\,.
$$
We present expressions for $E_k$ below and here note that not only
integrals of stationary equation \eqref{KK} but also the equation
itself enter the spectral curve equation:
$$
K=c\,u'+u^{{\sss(\text{V})}}+5\,u\,u'''+\frac{25}{2}\,u'\,u''+
5\,u^2\,u'\,.
$$
By virtue of  \eqref{KK} the constant $K$ must vanish identically. But
the expression $u''+2\,u^2+2\,c=\mbox{const}$ also does not contradict
it. If the constant in the right-hand side of this equation is nonzero,
then the compatibility with the equation $K = 0$ results in the trivial
answer $u=\mbox{const}$. Otherwise, we have the degenerate curve
$$
\mu^3 + 27\,c\,\lambda^2\,\mu+729\,\lambda^5-27\left(\frac34\,u'^2+
u^3+3\,c\,u\right)\!\lambda^3=0\,.
$$
This curve has the genus $g = 1$, the corresponding solution
$u=-3\,\wp\big(x+\frac34\,g_2^{}t\big)$ can be easily calculated, and
we are not interested in it. In the nondegenerate case, curve \eqref{W}
becomes
\begin{equation}\label{ML}
\mu^3 + (27\,c\,\lambda^2 + E_2)\,\mu + 729\,\lambda^5 +
81\,E_1\,\lambda^3 + E_3\,\lambda = 0\,,
\end{equation}
it has the genus $g = 4$, and the problem of determining the
corresponding potential is nontrivial.

\subsection{The rational trace formula}
We first obtain the standard trace formula, \ie, the representation of
the potential in form \eqref{R}. Let us determine the quantities
$\gamma_k^{}$. In the finite-gap potential theory, these objects are
the poles of $\Psi(x)$, which are independent of $\lambda$. Eliminating
the $\Psi$-function and its first derivative from Eqs.~\eqref{3} and
\eqref{ABC}, we obtain the expression
\begin{equation}\label{ln}
\frac{\Psi'}{\Psi}=27\,\lambda\,\frac{3\,\mu^2+Q}{\Pi}-
\frac{1}{18\,\lambda}
(u''+2\,u^2+2\,c)\,.
\end{equation}
Here $Q(\lambda)\DEF 27\,c\,\lambda^2+E_1([u])$,  and the differential
polynomial $\Pi=\Pi(\lambda;[u])$ is
\begin{equation}\label{Pi}
\begin{split}
3^{-7}\,\Pi(\lambda;[u])&=\lambda^4-\frac16\,u'\,\lambda^3+
\frac{1}{18}\big(
u^{{\sss(\text{IV})}}
+
5\,u\,u''+4\,u'^2+2\,u^3+2\,c\,u\big)\lambda^2+{}\\
&\quad+\frac{1}{324}(u'''+4\,u\,u')(u''+2\,u^2+2\,c)\,\lambda+
\frac{1}{3^6\,8}(u''+2\,u^2+2\,c)^3\,.
\end{split}
\end{equation}
We now present the Novikov integrals  $E_k$:
\begin{align}\label{E1}
E_1&=u^{{\sss(\text{IV})}}+5\,u\,u''+
\frac{15}{4}\,u'^2+\frac53\,u^3+c\,u\,,\\\notag
E_2&=-\frac14\,u'''\,(u'''+8\,u\,u')-\frac14\,u\,u''^2+
\frac{1}{24}\,u''\,(E_1+3\,u'^2-20\,u^3-12\,c\,u)+\cdots \,,\\\notag
E_3&=-\frac34\,u'''\,u'\,(u''+2\,u^2+2\,c)+\frac14\,u''^3+
\frac34\,(u^2+c)\,u''^2-
3\,u'^2\,u''+\cdots\,,
\end{align}
where we truncate the expressions for $E_2$, $E_3$ because we do not
need them in what follows.

The separation variables  $\gamma_k^{}$ determine the poles of
logarithmic derivative \eqref{ln}, \ie, they factor the polynomial
$\Pi$. We therefore define
\begin{equation}\label{gamma}
\Pi=3^7\,(\lambda-\gamma_1^{})(\lambda-\gamma_2^{})
(\lambda-\gamma_3^{})(\lambda-\gamma_4^{})\,.
\end{equation}
If we define the second coordinate $\mu_k^{}=\mu_k^{}(x)$ by the
formula
$$
\mu_k^3+Q\big(\gamma_k^{}(x)\big)\,\mu_k^{}+T\big(\gamma_k^{}(x)\big)
=0\,,
$$
then we obtain an analogue of the Dubrovin equations from  \eqref{ln}
in the case under consideration. Indeed, passing to the limit
$\lambda\to \gamma_k^{}$ in (8), we obtain
\begin{equation}\label{Q}
\frac{d\gamma_k^{}}{dx}=-\frac{\gamma_k^{}}{81}\,
\frac{3\,\mu_k^2+Q(\gamma_k^{})}
{\prod\limits_{j\ne k}(\gamma_k^{}-\gamma_j^{})}\,.
\end{equation}
Additional explanations can be found in \cite{br}, where the general
method can also be found for obtaining such equations and the second
coordinate of the zero of the polynomial  $\Pi$:
\begin{equation}\label{mu}
\mu_k^{}=3\,u\,\gamma_k^{}+
\frac{(u''+2\,u^2+2\,c)^2}{36\,\gamma_k^{}}\,.
\end{equation}
Comparing \eqref{Pi} and \eqref{gamma}, for instance, we obtain the
identities
\begin{equation}\label{ident}
u'=6\,\sideset{}{_k}\sum_1^4\gamma_k^{}\,,\qquad
u^{{\sss(\text{IV})}}+5\,u\,u''+4\,u'^2+2\,u^3+2\,c\,u=18\,
\sum_{k>j}\gamma_k^{}\,\gamma_j^{}\,.
\end{equation}
Composing symmetric combinations of Eqs.~\eqref{Q} and using formula
\eqref{mu} for the second coordinate, we can eliminate all the
derivatives of the potential and thus obtain a linear equation for $u$.
This is the desired trace formula. Omitting the details, we eventually
obtain
\begin{equation}\label{trace}
u=\frac13\,\sideset{}{_k}\sum_{1}^4\frac{\gamma_k^2\,\mu_k^{}}
{\prod\limits_{j\ne
k}(\gamma_k^{}-\gamma_j^{})}\,.
\end{equation}
The last simplification of Eqs.~\eqref{Q} is to reduce them to
quadratures. Again composing symmetric combinations of the right-hand
sides of  \eqref{Q}, we obtain the equations
$$
\begin{aligned}
\sideset{}{_k}\sum_{1}^4\frac{d\gamma_k^{}}{3\,\mu_k^2+
Q(\gamma_k^{})}&=0\,,\\
\sideset{}{_k}\sum_{1}^4\frac{\mu_k^{}\,d\gamma_k^{}}
{3\,\mu_k^2+Q(\gamma_k^{})}&=0\,,
\end{aligned}
\qquad\qquad
\begin{aligned}
\sideset{}{_k}\sum_{1}^4\frac{\gamma_k^{}\,d\gamma_k^{}}{3\,\mu_k^2+
Q(\gamma_k^{})}&=0\,,\\
\sideset{}{_k}\sum_{1}^4\frac{\gamma_k^2\,d\gamma_k^{}}{3\,\mu_k^2+
Q(\gamma_k^{})}&=
-\frac{1}{81}\,dx\,,
\end{aligned}
$$
which, as expected, result in a Jacobi problem, \ie, in the problem of
inverting four holomorphic integrals on curve  \eqref{ML}:
\begin{align*}
\sideset{}{_k}\sum_{1}^4 \!\!\int\limits_{}^{(\gamma_k^{},\,\mu_k^{})}
\!\!\!\!\!\frac{d\lambda} {3\,\mu^2+Q(\lambda)}&= a_1^{},&
\sideset{}{_k}\sum_{1}^4 \!\!\int\limits_{}^{(\gamma_k^{},\,\mu_k^{})}
\!\!\!\!\! \frac{\lambda\,d\lambda} {3\,\mu^2+Q(\lambda)}&=a_2^{},
\\
\sideset{}{_k}\sum_{1}^4 \!\!\int\limits^{(\gamma_k^{},\,
\mu_k^{})} \!\!\!\!\!
\frac{\mu\,d\lambda} {3\,\mu^2+Q(\lambda)}&=a_3^{}, &
\sideset{}{_k}\sum_{1}^4
\!\!\int\limits^{(\gamma_k^{},\,\mu_k^{})}
\!\!\!\!\!\frac{\lambda^2\,d\lambda}
{3\,\mu^2+Q(\lambda)}&=a_4^{}-\frac{1}{81}\,x\,.
\end{align*}
In what follows, we consider the dependence
$\gamma_k^{}=\gamma_k^{}(x)$ to be defined from these equations.

We note that this scheme for obtaining the Dubrovin equation and trace
formula from the definition of the differential polynomial  $\Pi$ and
$\Psi$ is intrinsic for general finite-gap operators. The definitions
of the quantities  $\gamma_k^{}$, the Dubrovin equations, and various
relations for the potential and its derivatives (different versions of
the trace formula) are ``simultaneously entangled'' in the general set
of determining polynomial relations.

\section{The transcendental trace formula}
\noindent Deriving the preceding formulas, we manipulated with
polynomials, which is common in polynomial ideal theory  \cite{cox}.
The bases of such ideals are not unique, which means that we can
presumably find other  relations determining the potential. This is
indeed the case. Eliminating the higher derivative
$u^{(\sss\text{IV})}$ from the Novikov integral  $E_1$ and from the
second identity in \eqref{E1}, we obtain the quadratic equation for
$u'$:
$$
3\,u'^2+4\,u^3+12\,c\,u+E_1=216\,\sum_{k>j}\gamma_k^{}\gamma_j^{}\,.
$$
But  $u'$ can be expressed uniquely in terms of  $\gamma$ by virtue of
the first identity in \eqref{ident}. We then obtain the cubic
polynomial in the variable $u$:
\begin{equation}\label{cubic}
u^3+3\,c\,u+3\,E_1+27\,\sideset{}{_k}\sum_1^4\gamma_k^2=0\,.
\end{equation}
It can be interpreted as an algebraic (implicit) variant of the
principal trace formula.

On the other hand, finite-gap potentials are single-valued functions of
the variable $x$. They satisfy autonomous ordinary differential
equations, pass the Painlev\'e test, and are expressed in terms of
theta functions. It is also known (but seldom used) that roots of any
3rd or 4th degree polynomial can be expressed analytically via its
coefficients in terms of elliptic functions  \cite{a}. For exhaustive
information on this topic see \cite{br3}. In our case the situation
simplifies since we may directly consider polynomial \eqref{cubic} as a
Weierstrass cubic polynomial
\begin{align*}
4\,u^3+12\,c\,u+12\,E_1+108\,\sideset{}{_k}\sum_1^4\gamma_k^2(x)&=
4\,u^3-a\,u-b={}\\
&=4\,(u-e)(u-\ep)(u-\epp)\,,
\end{align*}
where the points $e$, $\ep$, and $\epp$ depend on the coefficients $a$
and $b$, which in turn depend on $x$. The roots of this polynomial are
$$
u=\{e,\ep,\epp \}=\wp\big(\omega_k^{}(x);a,b\big)\,,
$$
where
$$
a=-12\,c\,,\qquad b=-108\textstyle\sum\gamma_k^2(x)-12\,E_1\,,
$$
are expressed in terms of the theta constants \cite{a}
\begin{alignat}{12}
e&=&\frac{1}{\omega^2}\,\frac{\pi^2}{12}
\big\{\vartheta_3^4(\tau)+\vartheta_4^4(\tau)
\big\}\,,\notag\\[0.3em]
\ep&=-&\frac{1}{\omega^2}\,\frac{\pi^2}{12}
\big\{\vartheta_2^4(\tau)+\vartheta_3^4(\tau)
\big\}\,,\label{e}\\[0.3em]
\epp&=&\frac{1}{\omega^2}\,\frac{\pi^2}{12}
\big\{\vartheta_2^4(\tau)-\vartheta_4^4(\tau) \big\}\,,\notag
\end{alignat}
and the quantities  $\omega$ and $\tau=\omega'/\omega$ are expressed in
terms of the coefficients $a$ and $b$. It suffices to take just one
such expression and express the $\vartheta$-constants in \eqref{e}
using the classic series  \cite{a}:
$$
\vartheta_2(\tau)\DEF\re^{\frac\pi4\ri\tau}_{\mathstrut}
\sideset{}{_k}\sum_{-\infty}^\infty\!
\re^{(k^2+k)\pi\ri\tau}_{\mathstrut},\qquad
\vartheta_3(\tau)\DEF
\sideset{}{_k}\sum_{-\infty}^\infty\!
\re^{k^2\pi\ri\tau}_{\mathstrut} ,\qquad
\vartheta_4(\tau)\DEF
\sideset{}{_k}\sum_{-\infty}^\infty\!(-1)^k
\re^{k^2\pi\ri\tau}_{\mathstrut} .
$$
(we let the symbol $\re^z$ denote the exponential function to
distinguish it from the Weierstrass $e$-points).

We thus obtain a classic elliptic modular function inversion problem,
\ie, the problem of expressing the periods $(2\,\omega,2\,\omega')$ of
the elliptic curve  $w^2=4\,z^3-a\,z-b$ in terms of its coefficients.
The solution scheme is known. We must find the root $\tau$ of the
transcendental equation
$$
J(\tau)=\frac{a^3}{a^3-27\,b^2}\,,
$$
where $J(\tau)$ is the classic modular Klein function \cite{a,bateman}.
After finding the root $\tau$, we calculate the half-periods
$(\omega$, $\omega')$ by the formula
\begin{equation}\label{om}
\omega^2=\frac{a}{b}\,\frac{g_3^{}(\tau)}{g_2^{}(\tau)}\,,\qquad
\omega'=\tau\,\omega\,,
\end{equation}
where  $g_{2,3}^{}(\tau)$ are the known modular forms  \cite{bateman}.
We have the Eisenstein series, the Hurwitz--Lambert series, or the
$\vartheta$-constant representations for these forms:
\begin{equation}\label{g23}
\begin{split}
g_2^{}(\tau)&=\frac{\pi^4}{24}\,\big\{\vartheta_2^8(\tau)+
\vartheta_3^8(\tau)+\vartheta_4^8(\tau) \big\}\\
g_3^{}(\tau)&=\frac{\pi^6}{432}\,\big\{\vartheta_2^4(\tau)+
\vartheta_3^4(\tau)\big\}\big\{\vartheta_3^4(\tau)+
\vartheta_4^4(\tau)\big\} \big\{\vartheta_4^4(\tau)-
\vartheta_2^4(\tau)\big\}\,.
\end{split}
\end{equation}

It remains to write the expression for the root $\tau$. Strangely
enough, an explicit analytic solution of this classic problem is
lacking\footnote{Unfortunately, solutions of this problem are given
incorrectly both on page 789 of encyclopedia  \cite{enc} and in the end
of Sect.~11 of the remarkable book \cite{a}; this is not due to
misprints.}, although it is common knowledge that this solution can be
written in terms of ratios of hypergeometric $_2F_1$-series. Because
the series  $_2F_1(J)$ converges only inside the unit circle, the
formulas differ depending on whether $|J| > 1$ or $|J| < 1$.
Altogether, this results in cumbersome expressions also containing
functional series in the logarithmic derivative of the Euler
$\Gamma$-function. For example, the corresponding expression (22)--(27)
in Sec.~14.6.2 of  \cite{bateman} is half a page long (even if we
forgive the incorrectness of formula (23)). But the problem admits a
simple solution.

We use the well-known fact that the function $J$ is closely related to
the hypergeometric equation of the form
$$
J(J-1)\,\psi''+\frac16\,(7\,J-4)\,\psi'+\frac{1}{144}\,\psi=0\,.
$$
Solutions of general hypergeometric equations are hypergeometric
functions  $_2F_1(a,b;c|z)$, but these functions can be rewritten in
terms of classic special functions under special restrictions imposed
on the parameters $a$, $b$, and $c$. We mean not reductions of the
function $_2F_1$ but the cases where the  $_2F_1$-series admits a
quadratic transformation, and the hypergeometric equation then reduces
to a two-parameter equation, \eg, to the Legendre equation. This is
true for the above equation. Its solution is a linear combination
$$
\psi=\sqrt[6]{J\,}\,\big\{A\,\mathrm{P}_\nu^\mu(\sqrt{1-J\,})+
B\,\mathrm{Q}_\nu^\mu(\sqrt{1-J\,})\big\}
$$
of the Legendre functions with the parameters
$(\nu,\,\mu)=\big({-}\frac12,\,\frac13\big)$ (see \cite{bateman2} for
detailed information about these functions). From the above, we
conclude that the formula
\begin{equation}\label{invers}
\tau=\frac{\boldsymbol{a}\,\mathrm{P}(\sqrt{1-J\,})+
\boldsymbol{b}\,\mathrm{Q}(\sqrt{1-J\,})}
{\boldsymbol{c}\,\mathrm{P}(\sqrt{1-J\,})+
\boldsymbol{d}\,\mathrm{Q}(\sqrt{1-J\,})}
\end{equation}
holds at certain numerical values of $\{\boldsymbol{a}, \boldsymbol{b},
\boldsymbol{c}, \boldsymbol{d}\}$ (we omit the indices  $\nu$, $\mu$ of
the Legendre functions for brevity). We present only the final answer;
see \cite{br3} for derivation.

\begin{proposition}
For the elliptic curve $w^2=4\,z^3-a\,z-b$, the period ratio
$\tau=\omega'/\omega$ is
\end{proposition}
\begin{equation}\label{PQ}
\tau=\mbig[9]\{\pi\,\ri\,\frac{\mathrm{P}
\big(\sqrt{1-\mathbf{J}\,}\big)}
{\mathrm{Q}\big(\sqrt{1-\mathbf{J}\,}\big)}-1
\mbig[9]\}\,\re^{\!\frac{\pi}{3}\ri}_{\mathstrut}\,,\qquad
\mbox{where}\quad \mathbf{J}\DEF \frac{a^3}{a^3-27\,b^2}\,.
\end{equation}

\begin{remark}
If  $\mathbf{J}$ is substituted for $J(\tau)$ in equality  \eqref{PQ},
then this equality becomes an identity that holds for all $\tau$ in the
upper half-plane  $\mathbb{H}^+$. The left- and right-hand sides of
this identity must be calculated up to the action of the total modular
group  $\mathrm{PSL}_2(\mathbb{Z})=:\boldsymbol{\Gamma}(1)$. Relation
\eqref{PQ} is therefore a  $\boldsymbol{\Gamma}(1)$-equivalent of the
solution of the modular inversion problem in the Legendre
representation. Namely, given the elliptic curve in the Legendre form
$w^2=(1-z^2)(1-k^2z^2)$, we can calculate the elliptic modulus  $\tau$
by the celebrated Jacobi formula
\begin{equation}\label{KK'}
\tau=\ri\,\frac{K'(k)}{K(k)}
\mod\;\boldsymbol{\Gamma}(2)\,,
\end{equation}
where $K$ and $K'$ are the complete Legendre elliptic integrals of the
first kind  \cite{a,bateman,WW}.
\end{remark}

Returning to the potential, we use formulas  \eqref{e} and \eqref{om}
with  \eqref{g23} taken into account to obtain the desired
transcendental ``trace formula'':
\begin{equation}\label{trans}
u(x)=\frac{3}{2\,c}\,\mbig[9]\{
E_1+9\,\sideset{}{_k}\sum_1^4\gamma_k^2(x)\mbig[9]\}
\,\frac{\vartheta_2^8(\tau)+
\vartheta_3^8(\tau)+\vartheta_4^8(\tau)}{\big\{ \vartheta_2^4(\tau)+
\vartheta_3^4(\tau)\big\}\big\{ \vartheta_4^4(\tau)-
\vartheta_2^4(\tau)\big\}}\,,
\end{equation}
where  $\tau$ is the function
\begin{equation}\label{J}
\tau(x)=\widehat{\boldsymbol{\Gamma}}
\mbig[9](\pi\,\ri\,\frac{\mathrm{P}\big(\sqrt{1-\mathbf{J}\,}\big)}
{\mathrm{Q}\big(\sqrt{1-\mathbf{J}\,}\big)}\,
\re^{\frac{\pi}{3}\ri}_{\mathstrut}-
\re^{\frac{\pi}{3}\ri}_{\mathstrut} \mbig[9])\,,\qquad
\mathbf{J}\DEF\frac{4\,c^3}
{4\,c^3+9\,\big\{E_1+9\,\sum\gamma_k^2(x)\big\}^2}\,,
\end{equation}
and we let the symbol  $\widehat{\boldsymbol{\Gamma}}$ denote the
operation of setting a number to the fundamental domain of the group.
Formulas \eqref{trans}--\eqref{J} do not contain the coordinates
$\mu_k^{}$. Verifying that function \eqref{trans} indeed satisfies
relation  \eqref{E1} or a stationary version of Eq.~\eqref{KK} is a
good and rather nontrivial exercise.

\section{Transcendental identities for $\Theta$-functions}
\noindent The results in the preceding section demonstrate that we can
analogously obtain many more examples of Abelian functions that are
radicals or roots of algebraic equations. For example, the last summand
in polynomial  \eqref{Pi} provides a single-valued representation of
the Abelian function radical of the form
$$
18\,\sqrt[\uproot{1}3]{\gamma_1^{}\gamma_2^{}\gamma_3^{}\gamma_4^{}}=
u''+2\,u^2+2\,c\,,
$$
and the general mechanism for constructing ``transcendental traces'' is
based on eliminating derivatives of the potential from the total set of
relations determining the integrals $E_k$ and the variables
$\gamma_k^{}$. Solving the obtained equations in single-valued
functions (which are always theta functions), we obtain the answer.
These equations (and even possibly their orders) are not uniquely
defined and may depend on the genus, but transitions between them are
just different representations of the same Abelian function. They
differ even in the $(\gamma,\mu)$-representation because the
coordinates $\gamma$ and $\mu$ are algebraically related.

Now equating the transcendental and rational representations, we obtain
an unusual identity containing theta functions. Indeed, comparing
Eqs.~\eqref{trace} and \eqref{trans}, for example, \eqref{trans}
\begin{equation}\label{tr}
\frac13\,\sideset{}{_k}\sum_1^4\frac{\gamma_k^2\,\mu_k^{}}
{\prod\limits_{j\ne
k}(\gamma_k^{}-\gamma_j^{})}= \frac{3}{2\,c}\,\mbig[9]\{
E_1+9\,\sideset{}{_k}\sum_1^4\gamma_k^2\mbig[9]\} \,
\frac{\vartheta_2^8(\tau)+
\vartheta_3^8(\tau)+\vartheta_4^8(\tau)}{\big\{ \vartheta_2^4(\tau)+
\vartheta_3^4(\tau)\big\}\big\{ \vartheta_4^4(\tau)-
\vartheta_2^4(\tau)\big\}}\,,
\end{equation}
we can in principle represent the Abelian functions rational in
$(\gamma,\mu)$ in this formula in the form of $\Theta$-function
relations  \cite{buch}. The obtained relation for  $\Theta$-functions
of genus $g = 4$ is a transcendental identity containing the theta
constants $\vartheta\big( \tau(x)\big)$, although curve  \eqref{ML},
generally speaking, is not related to any elliptic curve. In addition
to the curve moduli, this identity also contains an arbitrary parameter
$x$. We can also move all symmetric functions that are rational in
$(\gamma, \mu)$ to the left-hand side of the equality. We then obtain
an Abelian function for the Jacobian of trigonal curve  \eqref{ML} in
the left-hand side and an   ``exotic'' representation of this curve in
terms of the elliptic theta constants of the Legendre functions in the
right-hand side:
\begin{equation}\label{tr2}
u^2=-2\,c\, \frac{\big\{ \vartheta_2^4(\tau)+
\vartheta_3^4(\tau)\big\}\big\{ \vartheta_4^4(\tau)-
\vartheta_2^4(\tau)\big\}}{\vartheta_2^8(\tau)+
\vartheta_3^8(\tau)+\vartheta_4^8(\tau)}-3\,c\,.
\end{equation}
In turn, arguments of the Legendre functions (hypergeometric functions)
in this representation are expressions containing another Abelian
function via the quantity $\sqrt{1-\mathbf{J}}$, \ie,
$\sum\gamma_k^2(x)$, by virtue of formula \eqref{J}. This function is
represented by its own rational fraction of $\Theta$-functions of the
type $\Theta(x\boldsymbol{U}+\boldsymbol{D})$. The nature of
transcendental relations is therefore rather involved and still needs
to be understood. The transcendency is preserved even when the Jacobian
of curve  \eqref{ML} splits completely into elliptic curves, and the
multidimensional  $\Theta$-functions become the elliptic
$\theta$-functions. Polynomials of type  \eqref{cubic} then remain
nondegenerate, and it is natural to expect the appearance of
transcendental  $\theta$-identities.

The type of transcendency obtained above is related to the elliptic
modular inversion because this inversion is generated by a solution of
an algebraic equation with application of the elliptic modular
functions. The general theta functions and constants appear in
more-general methods (see, \eg, Umemura's contribution to
\cite{mumford}), and it is therefore natural to expect the appearance
of other types of transcendental identities.

In conclusion, we note that the simultaneous and transcendental
unification of theta functions and hypergeometric series is in fact not
unexpected, because a natural feature of $_2F_1$ series are their
monodromy groups, which are in turn automorphisms of analytic
automorphic functions. All such functions that are currently known are
ratios of theta constants. Informally speaking, the ``theta'' geometry
and ``hypergeometry'' are mutually inverse (they realize inversions of
fractions of type \eqref{invers}), and our transcendental identities
reflect both this fact and the properties of functions of the type
$\Theta(x\boldsymbol{U}+\boldsymbol{D})$. Namely, Abelian manifolds are
parameterized as algebraic manifolds by polynomials in theta functions
while linear sections of Jacobians appear from integrable equations. A
partial illustration is already provided by Jacobi equation \eqref{KK'}
rewritten in the form
$$
\tau\equiv\ri\,
\frac{K'\!\left(\frac{\vartheta_2^2(\tau)}{\vartheta_3^2(\tau)}
\right)}{K\!\left(\frac{\vartheta_2^2(\tau)}{\vartheta_3^2(\tau)}
\right)} \mod\; \boldsymbol{\Gamma}(2)\qquad\forall
\tau\in\mathbb{H}^+\,.
$$
Considering this equation in a sufficiently small vicinity of a point
$\tau_{\sss0}$, we can drop out a ``not completely analytic'' operation
of bringing a point into the fundamental domain of the group, and this
equation then becomes an exact analytic transcendental identity. We can
proceed in the same way with the theta versions of our relations
\eqref{tr} and \eqref{tr2} interpreting them as functions of $x$.

\end{document}